\documentstyle{aipproc}
\begin{document}
\title{Muon-proton Colliders: Leptoquarks, Contact Interactions and Extra
Dimensions
\footnote{Invited talk at the 5th International Conference on
the Physics Potential and Development of $\mu^+ \mu^-$ Colliders, 
San Francisco CA, December 1999. Work supported by DOE.}
}
\author{Kingman Cheung}
\address{
Department of Physics, University of California at Davis, 
Davis, CA 95616}
\maketitle
\begin{abstract}
We discuss the physics potential of the $\mu p$ collider; especially,
leptoquarks, leptogluons, $R$-parity violating squarks, contact
interactions, and large extra dimensions.  We calculate the
sensitivity reach for these new physics at $\mu p$ colliders of
various energies and luminosities.
\end{abstract}

\section*{Introduction}
The R\&D \cite{rd,fmc} of the muon collider is well underway.  The
First Muon Collider (FMC) will have a 200 GeV muon beam on a 200 GeV
anti-muon beam, which could possibly be at the Fermilab \cite{fmc}.
With the existing Tevatron proton beam the muon-proton collision
becomes a possible option.  It would be a 200 GeV $\otimes$ 1 TeV $\mu
p$ collider. The existing lepton-proton collider is the $ep$ collider
at HERA.  Lepton-proton colliders have been proved to be successful by
the physics results from HERA.  In this work, we shall discuss the
physics potential of the $\mu p$ colliders at various energies and
luminosities.  Other $\mu p$ colliders that we consider in this study
are summarized in Table \ref{table1}.  The nominal yearly luminosity
of the 200 GeV $\otimes$ 1 TeV $\mu p$ collider is about 13 fb$^{-1}$.
Luminosities for other designs are roughly scaled by one quarter power
of the muon beam energy and given in Table \ref{table1}.

\begin{table}[t!]
\caption{
The center-of-mass energies $\sqrt{s}$ and luminosities ${\cal L}$ 
for various designs of muon-proton colliders.}
\label{table1}
\begin{tabular*}{5.8in}{@{\extracolsep{1in}}ccc}
& $\sqrt{s}({\rm GeV})$  & ${\cal L}\;({\rm fb}^{-1})$ \\
\tableline
\tableline
$30{\rm GeV} \otimes 820{\rm GeV}$ &  314 & 0.1 \\
$50{\rm GeV} \otimes 1{\rm TeV}$ &    447 & 2 \\
$200{\rm GeV} \otimes 1{\rm TeV}$  &  894 & 13 \\
$1{\rm TeV} \otimes 1{\rm TeV}$ &     2000 & 110 \\
$2{\rm TeV} \otimes 3{\rm TeV}$  &    4899 & 280 \\
\end{tabular*}
\end{table}

\section*{Physics Potential}

The physics opportunities of $\mu p$ colliders are similar to
those of $ep$ colliders, but the sensitivity reach might be very different,
which depends on how precise the particles can be identified and measured
in $ep$ and $\mu p$ environments.  
Similar to $ep$ colliders the proton structure functions can be measured 
to very large $Q^2$ and small $x$ in $\mu p$ colliders of 
higher energies.  At the 
200 GeV $\otimes$ 1 TeV $\mu p$ collider the $Q^2$ can be measured up to 
$10^6$ GeV$^2$. In addition,  QCD studies, 
search for supersymmetry and other exotic 
particles can also be carried out.  Here we 
concentrate on leptoquarks, leptogluons, $R$-parity 
violating squarks, $\mu$-$q$ contact interactions, and the large extra
dimensions. The goal here is to estimate the sensitivity reach for these
new physics at various energies and luminosities.

\subsection*{Leptoquarks and Leptogluons}

The second generation leptoquarks made up of a muon and 
a charm or strange quark are particularly interesting at the $\mu p$
collider because they can be directly produced in the $s$-channel 
processes, 
$\mu^\pm  c (s) \to L_{\mu c}\;(L_{\mu s})$.
It is conventional to assume no inter-generational mixing in order to 
prevent the dangerous flavor-changing neutral currents.  
The production cross section
of the leptoquark in $\mu p$ collisions is
\begin{equation}
\sigma = \frac{\pi \lambda^2}{2 s} \; q(x,Q^2) \times (J+1) \;,
\end{equation}
where $\lambda$ is the coupling constant and $J$ is the spin of the leptoquark.

On the other hand, a leptogluon has a spin of either $1/2$ or $3/2$, 
a lepton quantum number (in this case it is the muon), and a color 
quantum number (the same as gluon.)
The interaction for a spin $1/2$ leptogluon is given by
\begin{equation}
{\cal L} = g_s \frac{M_{L_{\mu g}}}{2 \Lambda_{\mu g}^2} 
\overline{ L^a_{\mu g} }
\sigma^{\mu \nu} \mu \, G_{\mu \nu}^b \, \delta_{a b}  + {\rm h.c.} \;,
\end{equation} 
where $\Lambda_{\mu g}$ 
is the scale that determines the strength of the interaction.
The leptogluon can also be produced in the $s$-channel and 
the production cross section is
\begin{equation}
\sigma = \frac{4 \pi^2 \alpha_s}{s}\,
    \left( \frac{M^2_{L_{\mu g}}}{\Lambda_{\mu g}^2} \right )^2
    g(x,Q^2) \;,
\end{equation}
where $g(x,Q^2)$ is the gluon luminosity.

The $R$-parity violating squarks can be considered special scalar leptoquarks
that are the SUSY partners of quarks.
The cross section for $\mu^+ p \to \tilde{t}_L$ is given by
\begin{equation}
\label{90}
\sigma_{\tilde{t}_L} = \frac{\pi |\lambda'_{231}|^2}{4s} \; d \left(
\frac{m^2_{\tilde{t}_L}}{s}, Q^2=m_{\tilde{t}_L}^2 \right ) \;,
\end{equation}
where $d$ is the down-quark luminosity.  
The above formula can be easily modified to the production of
other squarks with the corresponding subscripts in $\lambda'$ and
parton functions. 

If kinematically allowed the leptoquarks, leptogluons, and the $R$-parity
violating squarks are produced in the $s$-channel and thus give rise to a
spectacular enhancement in a single bin of the invariant mass $M$ distribution
or the $x=M^2/s$ distribution.

\subsection*{Contact Interactions}
The effective four-fermion contact interactions can arise from 
fermion compositeness or exchanges of heavy particles like heavy $Z'$,
heavy leptoquarks, or other exotic particles.
The conventional Lagrangian for $ll q q$ ($l=e,\mu$) contact
interactions has the form \cite{me}
\begin{eqnarray}
L_{NC} &=& \sum_q \Bigl[ \eta_{LL}
\left(\overline{l_L} \gamma_\mu l_L\right)
\left(\overline{q_L} \gamma^\mu q_L \right)
+ \eta_{RR} \left(\overline{l_R}\gamma_\mu l_R\right)
                 \left(\overline{q_R}\gamma^\mu q_R\right) \nonumber\\
&& \qquad {}+ \eta_{LR} \left(\overline{l_L} \gamma_\mu l_L\right)
                             \left(\overline{q_R}\gamma^\mu q_R\right)
+ \eta_{RL} \left(\overline{l_R} \gamma_\mu l_R\right)
\left(\overline{q_L} \gamma^\mu q_L \right) \Bigr] \,, \label{effL}
\end{eqnarray}
where $\eta_{\alpha\beta}^{lq} = \epsilon 4\pi/{\Lambda^{lq}_{\alpha\beta}}^2$.
We introduce the reduced amplitudes 
$M^{\mu q}_{\alpha \beta}$, where the subscripts label the chiralities
of the initial lepton ($\alpha$) and quark ($\beta$). The SM tree-level reduced
amplitudes for $\mu q\to \mu q$ are
\begin{equation}
\label{reduced}
M^{\mu q}_{\alpha\beta}(\hat t) = -\frac{e^2 Q_q}{\hat t} +
\frac{e^2 }{\sin^2 \theta_{\rm w}  \cos^2 \theta_{\rm w}}
\frac{g_\alpha^\mu g_\beta^q}{\hat t - m_Z^2} \;,
\qquad \alpha,\beta = L,R \;.
\end{equation}
The new physics contributions to $M^{\mu q}_{\alpha\beta}$ from
the $\mu \mu qq$ contact interactions are 
$
\Delta M^{\mu q}_{\alpha \beta}  = \eta^{\mu q}_{\alpha \beta}
$.
The differential cross sections are given by \cite{me}
\begin{eqnarray}
{d\sigma(\mu^+p)\over dx\,dy} &=& {sx\over16\pi} \left\{ u(x,Q^2) \left[
\left| M_{LR}^{\mu u} \right|^2 + \left| M_{RL}^{\mu u} \right|^2 + (1-y)^2 \left(
\left| M_{LL}^{\mu u} \right|^2 + \left| M_{RR}^{\mu u} \right|^2 \right) \right]
\right. \nonumber\\
&&\quad\left. {}+ d(x,Q^2) \left[ \left| M_{LR}^{\mu d} \right|^2 + \left|
M_{RL}^{\mu d} \right|^2 + (1-y)^2 \left( \left| M_{LL}^{\mu d} \right|^2 + \left|
M_{RR}^{\mu d} \right|^2 \right) \right] \right\}  \label{bar-ep}\\
{d\sigma(\mu^-p)\over dx\,dy} &=& {sx\over16\pi} \left\{ u(x,Q^2) \left[ \left|
M_{LL}^{\mu u} \right|^2 + \left| M_{RR}^{\mu u} \right|^2 + (1-y)^2 \left( \left|
M_{LR}^{\mu u} \right|^2 + \left| M_{RL}^{\mu u} \right|^2 \right) \right] \right.
\nonumber\\
&&\quad\left. {}+ d(x,Q^2) \left[ \left| M_{LL}^{\mu d} \right|^2 + \left|
M_{RR}^{\mu d} \right|^2 + (1-y)^2 \left( \left| M_{LR}^{\mu d} \right|^2 + \left|
M_{RL}^{\mu d} \right|^2 \right) \right] \right\} \;.
\end{eqnarray}

\subsection*{Model of extra dimensions}

Arkani-Hamed, Dimopoulos and Dvali \cite{add} proposed that
in the extra dimensions gravity is free to propagate while the SM particles 
are restricted to a 3-D-brane.  The size of the extra 
dimensions is postulated to be as large as mm to solve the
hierarchy problem by bringing the effective Planck scale $M_S$ down to TeV.
It implies a new gravity interaction for the graviton in the bulk.
In our $3+1$ dimensional point of view, the graviton behaves as a tower of
closely-spaced Kaluza-Klein states. Each state still 
couples to the SM particles with a normal gravitational strength of order of
$1/M_{\rm Pl}$ but, however, there are a huge number of such states.  
Collectively, the overall coupling strength becomes of order of $1/M_S$.
In the presence of the new interaction the double differential cross section 
is given by \cite{cheung}.
\begin{eqnarray}
\frac{d^2\sigma (\mu^+ p)}{dx dy} &=& 
\frac{s x}{16\pi} \sum_q f_q(x)  \; \Biggr \{
(1-y)^2 ( |M_{LL}|^2 + |M_{RR}|^2)  + |M_{LR}|^2 + |M_{RL}|^2 \nonumber  \\
&&+
\frac{\pi^2}{2}\, (s x )^2 \, 
   \left( \frac{\cal F}{M_S^4} \right )^2 \; 
   (32 - 64 y +42 y^2 -10y^3 + y^4 ) \nonumber \\
&&+ 
2 \pi e^2 Q_e Q_q \;\left( \frac{\cal F}{M_S^4} \right )\;
  \frac{(2-y)^3}{y} \nonumber \\
&&+
\frac{2\pi e^2 }{\sin^2\theta_{\rm w} \cos^2\theta_{\rm w}}\;
  \left( \frac{\cal F}{M_S^4} \right )\;
  (s x)\; \frac{1}{-Q^2  - M_Z^2} \; \biggr[
g_a^e g_a^q (6y -6y^2 +y^3 ) \nonumber \\
&& + g_v^e g_v^q (y-2)^3  \biggr ] \Biggr \} 
\nonumber \\
&+&
\frac{\pi}{2}\, f_{g}(x) \, (sx)^3 \; 
   \left( \frac{\cal F}{M_S^4} \right )^2 \; 
 \; (1-y)(y^2 -2y + 2) \;.
\end{eqnarray}

Unlike the leptoquarks, the contact interactions and the extra dimensions 
do not enhance the cross section in a single invariant mass bin, instead,
they enhance the cross section at large $Q^2$.

\section*{Sensitivity Reach}

\begin{table}[t!]
\caption{\label{table2}
The 95\% sensitivity reach on $\Lambda_{\alpha\beta}^{\mu q},\;
(\alpha,\beta=L,R ;\, q=u,d)$ at various $\mu^+(\mu^-) p$ colliders. 
}
\begin{tabular}{c@{\extracolsep{-0.4in}}||cc|cc|cc|cc|cc}
\multicolumn{11}{c}{$\mu^+ p$} \\
\tableline
&
\multicolumn{2}{|c@{\extracolsep{-0.1in}}|}{$30{\rm GeV} \otimes 820{\rm GeV}$} &
\multicolumn{2}{|c@{\extracolsep{-0.1in}}|}{$50{\rm GeV} \otimes 1{\rm TeV}$} &
\multicolumn{2}{|c@{\extracolsep{-0.1in}}|}{$200{\rm GeV} \otimes 1{\rm TeV}$}  &
\multicolumn{2}{|c@{\extracolsep{-0.1in}}|}{$1{\rm TeV} \otimes 1{\rm TeV}$} &   
\multicolumn{2}{|c@{\extracolsep{-0.1in}}}{$2{\rm TeV} \otimes 3{\rm TeV}$}  \\
\tableline
$\sqrt{s}({\rm GeV})$ & \multicolumn{2}{|c|}{314} & 
                        \multicolumn{2}{|c|}{447} & 
                        \multicolumn{2}{|c|}{894} & 
                        \multicolumn{2}{|c|}{2000} & 
                        \multicolumn{2}{|c}{4899} \\
${\cal L}\;({\rm fb}^{-1})$ &\multicolumn{2}{|c|}{0.1} & 
                            \multicolumn{2}{|c|}{2}   &
                            \multicolumn{2}{|c|}{13}   &
                            \multicolumn{2}{|c|}{110}   &
                            \multicolumn{2}{|c}{280}    \\
\tableline
\tableline
 & $+$ & $-$ &  $+$ & $-$ &  $+$ & $-$ &  $+$ & $-$ & $+$ & $-$\\
$\Lambda_{LL}^{\mu u}$ & 
3.4  & 3.4     &  9.6 & 9.1  & 22.8  & 21.5   & 57.4 & 56.3 & 112.7 & 109.7 \\
$\Lambda_{LR}^{\mu u}$ & 
4.7  & 4.2     &  11.4 & 10.7  & 24.0 & 23.1   & 58.9 & 55.8 & 115.2 & 105.6 \\
$\Lambda_{RL}^{\mu u}$ & 
4.4  & 3.4     &  9.7 & 8.8  & 19.2 & 16.8   & 43.8 & 38.1 & 86.0 & 64.9  \\
$\Lambda_{RR}^{\mu u}$ & 
3.2  & 3.0     &  9.0 & 7.9  & 20.1 & 18.8   & 48.6 & 49.3 & 98.7 & 92.3 \\
\tableline
$\Lambda_{LL}^{\mu d}$ & 
   &      &  7.0 & 7.0     & 17.3 & 18.0   & 45.0 & 48.3  &  88.9 & 96.1  \\
$\Lambda_{LR}^{\mu d}$ & 
1.9 & 2.9     &  5.1 & 6.5     & 11.1 & 13.5   & 26.8 & 31.9  & 46.7  & 63.4 \\
$\Lambda_{RL}^{\mu d}$ & 
2.1 & 2.2     &  4.7 & 3.8     & 11.4 & 6.9   & 30.4 & 22.8  & 64.0 & 45.5  \\
$\Lambda_{RR}^{\mu d}$ & 
1.4 & 2.3     &  5.0 & 5.7     & 12.0 & 13.1   & 31.4 & 32.1  & 58.2 & 65.5 \\
$\Lambda_{VV}  $ & 
6.5   & 6.1     &  16.6 & 15.5  & 35.2 & 34.0 & 85.4 & 84.9 & 166.8 & 161.9 \\
$\Lambda_{AA}$   & 
2.8   & 5.0     &  10.6 & 11.8  & 22.6 & 25.8 & 58.6 & 62.3 & 109.4 & 125.0 \\
\tableline
\tableline
\multicolumn{11}{c}{$\mu^- p$} \\
\tableline
\tableline
 & $+$ & $-$ &  $+$ & $-$ &  $+$ & $-$ &  $+$ & $-$ & $+$ & $-$\\
$\Lambda_{LL}^{\mu u}$ & 
5.2  & 5.0 & 13.2 & 12.9  & 29.6  & 29.6 & 76.9 & 72.9 & 147.6 & 144.8 \\
$\Lambda_{LR}^{\mu u}$ & 
3.1  & 2.7 & 7.1 & 6.9  & 14.3 & 13.9 & 34.4 & 31.2 & 64.9 & 58.9 \\
$\Lambda_{RL}^{\mu u}$ & 
3.0  & 2.5 &  6.7 & 6.1  & 12.3 & 11.6 & 26.9 & 22.5 & 50.6 & 39.7  \\
$\Lambda_{RR}^{\mu u}$ & 
4.8  & 4.5 &  12.0 & 11.3  & 26.0 & 25.5 & 65.5 & 63.2 & 128.5 & 121.7 \\
\tableline
$\Lambda_{LL}^{\mu d}$ & 
2.9 & 3.4  &  8.2 & 8.5  & 19.3 & 20.5 & 50.0 & 53.3  & 101.1 & 102.4  \\
$\Lambda_{LR}^{\mu d}$ & 
1.6 & 2.2  &  4.2 & 4.7  & 8.8 & 9.6 & 19.4 & 22.9  & 38.6 & 44.6 \\
$\Lambda_{RL}^{\mu d}$ & 
1.6 & 1.9  &  3.8 & 3.3  & 7.1 & 4.9 & 20.5 & 14.3  & 45.7 & 37.7  \\
$\Lambda_{RR}^{\mu d}$ & 
2.1 & 2.8  & 6.0 & 6.6 & 13.4 & 14.6 & 33.3 & 37.0  & 64.3 & 71.7 \\
$\Lambda_{VV}  $ & 
6.5   & 6.4  &  16.4 & 15.6  & 35.7 & 34.4 & 87.7 & 85.9 & 173.7 & 162.9 \\
$\Lambda_{AA}$   & 
5.4   & 1.8  &  13.2 & 12.3  & 29.2 & 27.6 & 73.5 & 69.7 & 142.9 & 135.7 \\
\end{tabular}
\end{table}

The 95\% sensitivity reach on the contact interactions and extra dimensions
are calculated as follows.  
We use the the 2-dimensional $x$-$y$ distribution to calculate
the sensitivity to these new interactions, so as to
maximize the sensitivity \cite{greg}.
We divide the $x$-$y$ plane ($0.05<x<0.95$ and $0.05<y<0.95$) into a grid.
We calculate the number of events predicted by the standard model in each 
bin with an efficiency of 0.8.  We then 
follow the Monte Carlo approach in Ref. \cite{greg}.

The sensitivity reach on the contact interaction scales 
$\Lambda^{\mu q}_{\alpha \beta}$ is tabulated in
Table \ref{table2}.  The maximum reach of $\Lambda$ at each center-of-mass 
energy roughly scales as $\Lambda \sim 40 \sqrt{s}$.
The effect of luminosity on $\Lambda$ is rather small: $\Lambda$ only 
scales as the $1/4$th power of the luminosity.
The sensitivity reach on the effective Planck scale $M_S$ for the model
of large extra dimensions is tabulated in Table \ref{ms}.

\begin{table}[t!]
\caption{\label{ms}
The 95\% sensitivity reach on $\eta={\cal F}/M_S^4$ and the corresponding 
$M_S$ for $n=3-6$ at various $\mu^+(\mu^-) p$ colliders.
}
\begin{tabular}{c||c|c|c|c|c}
\multicolumn{6}{c}{$\mu^+ p$} \\
\tableline
 &
$30{\rm GeV} \otimes$  &
$50{\rm GeV} \otimes$  &
$200{\rm GeV} \otimes$   &
$1{\rm TeV} \otimes$   &   
$2{\rm TeV} \otimes$  \\
& $820{\rm GeV}$ &  $1{\rm TeV}$ & $1{\rm TeV}$ & $1{\rm TeV}$ & $3{\rm TeV}$\\
\tableline
$\sqrt{s}({\rm GeV})$ & {314} & 447 & 894 & 2000 & 4899 \\
${\cal L}\;({\rm fb}^{-1})$ & {0.1} & 2 & 13 & 110 & 280   \\
\tableline
\tableline
$\eta$ (TeV$^{-4}$)& 
2.24  & $1.69\cdot 10^{-1}$ & $9.35\cdot10^{-3}$ & $3.01\cdot 10^{-4}$  & 
  $1.38\cdot 10^{-5}$  \\
$M_S$ (TeV) & & & & & \\
$n=3$ &  
0.97 & 1.86 & 3.82 & 9.03  & 19.5 \\
$n=4$ & 
0.82 & 1.56 & 3.22 & 7.59  & 16.4 \\
$n=5$ & 
0.74  & 1.41 & 2.91 & 6.86  & 14.8 \\
$n=6$ &
0.69  & 1.31 & 2.70 & 6.38  & 13.8 \\
\tableline
\tableline
\multicolumn{6}{c}{$\mu^- p$} \\
\tableline
\tableline
$\eta$ (TeV$^{-4}$)& 
2.14  & $1.72\cdot 10^{-1}$ & $8.95\cdot 10^{-3}$ & $2.99\cdot 10^{-4}$  & 
 $1.38\cdot 10^{-5}$ \\
$M_S$ (TeV)& & & & & \\
$n=3$ &  
0.98  & 1.85 & 3.87 & 9.05  & 19.5 \\
$n=4$ & 
0.83  & 1.55 & 3.25 & 7.61  & 16.4 \\
$n=5$ & 
0.75  & 1.40 & 2.94 & 6.87 & 14.8\\
$n=6$ &
0.70  & 1.31 & 2.73 & 6.40 & 13.8
\end{tabular}
\end{table}

\begin{table}[t!]
\caption{\label{table3}
95\% sensitivity reach on $\lambda'_{231}\;(\lambda'_{213})$ for a few choices 
of $m_{\tilde{t}_L}\;(m_{\tilde{b}_R})$ at various $\mu^+ p$ 
$(\mu^- p)$ 
colliders. The subprocess is $\mu^+ d \to \tilde{t}_L$ ($\mu^- u \to 
\tilde{b}_R$). }
\begin{tabular}{c@{\extracolsep{-0.2in}}||c@{\extracolsep{-0.1in}}|c
@{\extracolsep{-0.1in}}|c@{\extracolsep{-0.1in}}|c@{\extracolsep{-0.1in}}|c
@{\extracolsep{-0.1in}}}
&
$30{\rm GeV} \otimes$  &
$50{\rm GeV} \otimes$  &
$200{\rm GeV} \otimes$   &
$1{\rm TeV} \otimes$   &   
$2{\rm TeV} \otimes$  \\
& $820{\rm GeV}$ &  $1{\rm TeV}$ & $1{\rm TeV}$ & $1{\rm TeV}$ & $3{\rm TeV}$\\
\tableline
$\sqrt{s} ({\rm GeV})$ & 314 & 447  & 894 & 2000 & 4899 \\
\hline
${\cal L} ({\rm fb}^{-1})$ & 0.1 & 2 & 13 & 110 & 280 \\
\tableline 
\tableline
$m_{\tilde{t}_L,\tilde{b}_R} \; ({\rm GeV})$ & & & & & \\
200 & 0.014 (0.0097) & 0.0043 (0.0036) & 0.0025 (0.0023)& 0.0015 (0.0014)& 
  0.0010 (0.0010)\\ 
300 & $\infty$ (0.23)& 0.0091 (0.0062) & 0.0031 (0.0028)& 0.0019 (0.0018)& 
  0.0014 (0.0014)\\
400 &  -     & 0.062 (0.029) & 0.0039 (0.0034)& 0.0021 (0.0020)& 
  0.0017 (0.0016)\\
500 &  -    &   -  & 0.0054 (0.0042)& 0.0024 (0.0022)& 0.0019 (0.0019)\\
600 &  -    &   -  & 0.0083 (0.0057)& 0.0026 (0.0024)& 0.0021 (0.0020)\\
700 &  -    &   -  & 0.016 (0.0095)& 0.0029 (0.0026)& 0.0023 (0.0022)\\
800 &  -    &   -  & 0.067 (0.027)& 0.0032 (0.0028)& 0.0025  (0.0023)\\
900 &  -    &   -    &   -   & 0.0036 (0.0031)& 0.0026 (0.0024) \\
1000 & -    &   -    &   -   & 0.0041 (0.0034)& 0.0027 (0.0026) \\
1500 & -    &   -    &   -   & 0.012  (0.0071)& 0.0033 (0.0030) \\
2000 & -    &   -    &   -   &   -    & 0.0041 (0.0036) \\
2500 & -    &   -    &   -   &   -    & 0.0053 (0.0043) \\
3000 & -    &   -    &   -   &   -    & 0.0075 (0.0055)\\
3500 & -    &   -    &   -   &   -    & 0.012 (0.0078)\\
4000 & -    &   -    &   -   &   -    & 0.027 (0.014) \\
4500 & -    &   -    &   -   &   -    & 0.12 (0.052)\\
\end{tabular}
\end{table}

To estimate the sensitivity reach for $R$-parity violating squarks, leptoquarks
and leptogluons with a mass $m$, 
we assume the enhancement in cross section is in 
the mass bin of $(0.9m,\;  1.1m)$.  We calculate the number of 
events predicted by the standard model in this bin with an efficiency of 0.8, 
call it $n^{\rm sm}$.  Then we use the poisson statistics to
estimate the $n^{\rm th}$ that $n^{\rm sm}$ can fluctuate to at the 95\% CL.
Once the $n^{\rm th}$ is obtained the coupling constant $\lambda$ or
the leptogluon scale $\Lambda_{\mu g}$ can be determined.
These results are tabulated in Tables \ref{table3} to \ref{table5}.

\begin{table}[t!]
\caption{\label{table4}
95\% sensitivity reach on $\lambda^0\;\lambda^1$ for a few choices 
of $m_{L^0,L^1}$ at various $\mu^- p$ colliders. The subprocess is 
$\mu^- (c,s) \to L^{0,1}$. }
\begin{tabular}{c@{\extracolsep{-0.2in}}||c@{\extracolsep{-0.2in}}|
c@{\extracolsep{-0.2in}}|c@{\extracolsep{-0.2in}}|c@{\extracolsep{-0.2in}}|
c@{\extracolsep{-0.2in}}}
&
$30{\rm GeV} \otimes$  &
$50{\rm GeV} \otimes$  &
$200{\rm GeV} \otimes$   &
$1{\rm TeV} \otimes$   &   
$2{\rm TeV} \otimes$  \\
& $820{\rm GeV}$ &  $1{\rm TeV}$ & $1{\rm TeV}$ & $1{\rm TeV}$ & $3{\rm TeV}$\\
\tableline
$\sqrt{s} ({\rm GeV})$ & 314 & 447  & 894 & 2000 & 4899 \\
\hline
${\cal L} ({\rm fb}^{-1})$ & 0.1 & 2 & 13 & 110 & 280 \\
\tableline 
\tableline
\multicolumn{6}{c}{$\mu^- c (s) \to L^0$} \\
\tableline
$m_{L^0} \; ({\rm GeV})$ & & & & & \\
200 & 0.097 (0.072) & 0.017 (0.012) & 0.0040 (0.0033)& 0.0014 (0.0013)& 
  0.0008 (0.0008)\\ 
300 & 2.3 (2.3) & 0.071 (0.054) & 0.0081 (0.0062)& 0.0022 (0.0020)& 
  0.0012 (0.0011)\\
400 &  -     & 0.43 (0.43) & 0.016 (0.012)& 0.0031 (0.0026)& 
  0.0015 (0.0014)\\
500 &  -    &   -  & 0.031 (0.023)& 0.0043 (0.0035)& 0.0018 (0.0017)\\
600 &  -    &   -  & 0.065 (0.050)& 0.0059 (0.0047)& 0.0022 (0.0020)\\
700 &  -    &   -  & 0.15 (0.14)& 0.0079 (0.0061)& 0.0026 (0.0023)\\
800 &  -    &   -  & 0.38 (0.38)& 0.011 (0.0081)& 0.0030  (0.0027)\\
900 &  -    &   -    &   -   & 0.014 (0.011)& 0.0035 (0.0030) \\
1000 & -    &   -    &   -   & 0.019 (0.014)& 0.0040 (0.0034) \\
1500 & -    &   -    &   -   & 0.10  (0.090)& 0.0076 (0.0061) \\
2000 & -    &   -    &   -   &   -    & 0.014 (0.011) \\
2500 & -    &   -    &   -   &   -    & 0.026 (0.019) \\
3000 & -    &   -    &   -   &   -    & 0.049 (0.038)\\
3500 & -    &   -    &   -   &   -    & 0.10 (0.084)\\
4000 & -    &   -    &   -   &   -    & 0.22 (0.22) \\
4500 & -    &   -    &   -   &   -    & 0.57 (0.57)\\
\tableline
\tableline
\multicolumn{6}{c}{$\mu^- c (s) \to L^1$} \\
\tableline
$m_{L^1} \; ({\rm GeV})$ & & & & & \\
200 & 0.068 (0.051) & 0.012 (0.0087) & 0.0029 (0.0024)& 0.0010 (0.0009)& 
  0.0006 (0.0005)\\ 
300 & 1.6 (1.6) & 0.050 (0.038) & 0.0057 (0.0044)& 0.0016 (0.0014)& 
  0.0008 (0.0008)\\
400 &  -     & 0.30 (0.30) & 0.011 (0.0082)& 0.0022 (0.0019)& 
  0.0011 (0.0010)\\
500 &  -    &   -  & 0.022 (0.016)& 0.0031 (0.0025)& 0.0013 (0.0012)\\
600 &  -    &   -  & 0.046 (0.035)& 0.0042 (0.0033)& 0.0016 (0.0014)\\
700 &  -    &   -  & 0.11 (0.098)& 0.0056 (0.0043)& 0.0018 (0.0016)\\
800 &  -    &   -  & 0.27 (0.27)& 0.0075 (0.0057)& 0.0021  (0.0019)\\
900 &  -    &   -    &   -   & 0.010 (0.0076)& 0.0025 (0.0021) \\
1000 & -    &   -    &   -   & 0.014 (0.010)& 0.0029 (0.0024) \\
1500 & -    &   -    &   -   & 0.074  (0.063)& 0.0054 (0.0043) \\
2000 & -    &   -    &   -   &   -    & 0.0099 (0.0075) \\
2500 & -    &   -    &   -   &   -    & 0.018 (0.014) \\
3000 & -    &   -    &   -   &   -    & 0.035 (0.027)\\
3500 & -    &   -    &   -   &   -    & 0.072 (0.059)\\
4000 & -    &   -    &   -   &   -    & 0.16 (0.15) \\
4500 & -    &   -    &   -   &   -    & 0.40 (0.40)\\
\end{tabular}
\end{table}

\begin{table}[t!]
\caption{\label{table5}
95\% sensitivity reach on $\Lambda_{\mu g}$ for a few choices 
of $m_{L_{\mu g}}$ at various $\mu^- p$ colliders. The subprocess is 
$\mu^- g \to L_{\mu g}$. }
\begin{tabular}{c@{\extracolsep{-0.2in}}||c@{\extracolsep{-0.2in}}|c
@{\extracolsep{-0.2in}}|c@{\extracolsep{-0.2in}}|c@{\extracolsep{-0.2in}}|
c@{\extracolsep{-0.2in}}}
&$30{\rm GeV} \otimes 820{\rm GeV}$ & $50{\rm GeV} \otimes 1{\rm TeV} $
&$200{\rm GeV} \otimes 1{\rm TeV}$  & $1{\rm TeV} \otimes 1{\rm TeV} $
&$2{\rm TeV} \otimes 3{\rm TeV}$ \\
\tableline
$\sqrt{s} ({\rm GeV})$ & 314 & 447  & 894 & 2000 & 4899 \\
\hline
${\cal L} ({\rm fb}^{-1})$ & 0.1 & 2 & 13 & 110 & 280 \\
\tableline 
\tableline
\multicolumn{6}{c}{$\mu^- g \to L_{\mu g}\;\; (\Lambda_{\mu g})$ in TeV} \\
\tableline
$m_{L^0} \; ({\rm GeV})$ & & & & & \\
200 & 1.9 & 4.3 & 8.5 & 14.4 & 19.5 \\ 
300 & 0.2 & 3.3 & 9.0 & 17.0 & 23.9 \\
400 &  -  & 1.2 & 8.6 & 18.6 & 27.4 \\
500 &  -  &  -  & 7.8 & 19.6 & 30.4 \\
600 &  -  &  -  & 6.5 & 20.1 & 32.8 \\
700 &  -  &  -  & 4.7 & 20.2 & 35.0 \\
800 &  -  &  -  & 2.3 & 20.0 & 36.7 \\
900 &  -  &  -  &  -  & 19.5 & 38.2 \\
1000 & -  &  -  &  -  & 18.7 & 39.4 \\
1500 & -  &  -  &  -  & 11.9 & 42.4 \\
2000 & -  &  -  &  -  &   -  & 41.8 \\
2500 & -  &  -  &  -  &   -  & 38.8 \\
3000 & -  &  -  &  -  &   -  & 34.0 \\
3500 & -  &  -  &  -  &   -  & 27.2 \\
4000 & -  &  -  &  -  &   -  & 18.3 \\
4500 & -  &  -  &  -  &   -  & 7.7  \\
\tableline
\end{tabular}
\end{table}

\end{document}